\documentclass[twocolumn,showpacs,preprintnumbers,amsmath,amssymb]{revtex4}

\usepackage{graphicx}
\usepackage{dcolumn}
\usepackage{bm}

\begin{document}

\preprint{APS/123-QED}

\title{(La$_{1-x}$Ba$_x$)(Zn$_{1-x}$Mn$_x$)AsO: A Two Dimensional ``1111" Diluted Magnetic Semiconductor in Bulk Form}

\author{Cui Ding$^{1}$, Huiyuan Man$^{1}$, Chuan Qin$^{1}$, Jicai Lu$^{1}$, Yunlei Sun$^{1}$, Quan Wang$^{1}$, Biqiong Yu$^{1}$, Chunmu Feng$^{1}$, T. Goko$^{2}$, C.J. Arguello$^{2}$, L. Liu$^{2}$, B.J. Frandsen$^{2}$, Y.J. Uemura$^{2}$, Hangdong Wang$^{3}$, H. Luetkens$^{4}$, E. Morenzoni$^{4}$, W. Han$^{5}$, C. Q. Jin$^{5}$, T. Munsie$^{6}$, T.J. Williams$^{6}$, R.M. D'Ortenzio$^{6}$, T. Medina$^{6}$, G.M. Luke$^{6}$$^,$$^{7}$, T. Imai$^{6}$$^,$$^{7}$, F.L.
Ning$^{1,}$}\email{ningfl@zju.edu.cn}

\affiliation{$^{1}$Department of Physics, Zhejiang University,
Hangzhou 310027, China} \affiliation{$^{2}$Department of Physics,
Columbia University, New York, New York 10027, USA}
\affiliation{$^{3}$Department of Physics, Hangzhou Normal
University, Hangzhou 310016, China} \affiliation{$^{4}$Paul Scherrer
Institute, Laboratory for Muon Spin Spectroscopy, CH-5232 Villigen
PSI, Switzerland}\affiliation{$^{5}$Beijing National Laboratory for
Condensed Matter Physics, and Institute for Physics, Chinese Academy
of Sciences, Beijing 100190, China}\affiliation{$^{6}$Department of
Physics and Astronomy, McMaster University, Hamilton, Ontario
L8S4M1, Canada}\affiliation{$^{7}$Canadian Institute for Advanced
Research, Toronto, Ontario M5G1Z8, Canada}

\date{\today}


\begin{abstract}
We report the synthesis and characterization of a bulk diluted
magnetic semiconductor (La$_{1-x}$Ba$_x$)(Zn$_{1-x}$Mn$_x$)AsO (0
$\leqslant$ $x$ $\leqslant$ 0.2) with a layered crystal structure
identical to that of the ``1111" FeAs superconductors. No
ferromagnetic order occurs for (Zn,Mn) substitution in the parent
compound LaZnAsO without charge doping. Together with carrier doping
via (La,Ba) substitution, a small amount of Mn substituting for Zn
results in ferromagnetic order with $T_{C}$ up to $\sim$40 K,
although the system remains semiconducting. Muon spin relaxation
measurements confirm the development of ferromagnetic order in the
entire volume, with the relationship between the internal field and
$T_C$ consistent with the trend found in (Ga,Mn)As, the ``111"
Li(Zn,Mn)As, and the ``122" (Ba,K)(Zn,Mn)$_{2}$As$_{2}$ systems.
\end{abstract}


\maketitle


The successful fabrication of III-V ferromagnetic semiconductors
(Ga,Mn)As \cite{Ohno} in thin-film form via molecular beam epitaxy
(MBE) has triggered extensive research into diluted magnetic
semiconductors (DMS) \cite{Jungwirth,Dietl1,Zutic}. The highest
Curie temperatures, $T_C$, has been reported as $\sim$190 K with Mn
doping levels at $\sim$12 $\%$ in (Ga,Mn)As \cite{Wang}. However,
the quality of some thin films is strongly dependent on the
preparation procedure and heat treatment, and thus not always
controllable \cite{Samarth,Chambers}. Nonetheless, if properly
prepared, (Ga,Mn)As thin films exhibit spatially homogenous
ferromagnetism throughout the entire sample volume for a wide range
of Mn concentrations, as confirmed by muon spin relaxation ($\mu$SR)
measurements \cite{Dunsiger}. In contrast to the successful use of
MBE, the preparation of bulk (Ga,Mn)As has been much more
challenging, since the valence mismatch of Mn$^{2+}$ atoms and
Ga$^{3+}$ atoms results in severely limited chemical solubility,
i.e., $<$1$\%$. A similar dilemma was encountered in diluted
magnetic oxides (DMO) such as Co-doped ZnO and TiO$_2$, where
ferromagnetism has been observed in thin films but not in bulk
materials. The origin of ferromagnetism \cite{Chambers} in these DMO
systems is yet to be established.

Seeking for bulk DMS or DMO materials and the investigation of their
physical properties may shed light on the origin of ferromagnetism
in their thin film counterparts. Furthermore, the availability of
bulk specimens would enable the use of conventional magnetic probes
such as nuclear magnetic resonance (NMR) and neutron scattering, to
provide complementary information at a microscopic level. The
synthesis of bulk DMS or DMO specimens therefore becomes necessary.
Accordingly, Deng et al. \cite{Deng} followed a theoretical proposal
by Masek et al. \cite{Masek} and doped Mn into the direct-gap
semiconductor LiZnAs, thereby successfully synthesizing a bulk
I-II-V DMS system, Li(Zn,Mn)As, with ferromagnetic $T_C$ up to 50 K
\cite{Deng} and a cubic crystal structure not identical but very
similar to that of the ``111" LiFeAs \cite{Wangxc} and NaFeAs
\cite{Parker} superconductors. Li(Zn,Mn)As exhibits a very small
coercive field of 50 Oe, similar to that of (Ga,Mn)As.

Additionally, Deng et al.\cite{Deng} demonstrated that hole carriers
mediate ferromagnetism in both Li(Zn,Mn)As and (Ga,Mn)As with
exchange interactions of comparable magnitude. The growth of
Li(Zn,Mn)As compounds, however, suffers from difficulties in the
precise control of Li concentrations, making it difficult to
understand the interplay between charge carriers and spins
\cite{Deng}. Recently, Zhao et al.\cite{Zhao} reported another
ferromagnetic DMS system, (Ba,K)(Zn,Mn)$_{2}$As$_{2}$, with $T_C$ up
to $\sim$200 K and a crystal structure identical to that of the
``122" (Ba,K)Fe$_{2}$As$_{2}$ superconductors \cite{Johnston}. In
this letter, we report the successful fabrication of a new bulk DMS
material, (La$_{1-x}$Ba$_x$)(Zn$_{1-x}$Mn$_x$)AsO, with $T_C$ up to
40 K and a crystal structure identical to that of the ``1111" FeAs
superconductor LaFeAsO \cite{Kamihara}. This constitutes the third
example of a bulk DMS system structurally related to a family of
FeAs superconductors.

Chemically stable Ba and Mn atoms are doped into the parent
direct-gap ($\sim$1.5 eV) semiconductor LaZnAsO \cite{Kayanuma} to
introduce charge carriers and spins, respectively. The system
remains paramagnetic for Mn concentrations up to 10$\%$ in the
absence of Ba doping, but with doped charge carriers to mediate
magnetic exchange, static ferromagnetic order develops in
(La$_{1-x}$Ba$_x$)(Zn$_{1-x}$Mn$_x$)AsO below $T_C \sim 40$ K as
confirmed microscopically by $\mu$SR. Semiconducting behavior exists
for concentrations up to $x$ = 0.20, and pronounced magnetic
hysteresis with a coercive field of $\sim$1 T has also been
observed. Despite the striking difference between the
two-dimensional (2D) character of
(La$_{1-x}$Ba$_x$)(Zn$_{1-x}$Mn$_x$)AsO and the three-dimensional
(3D) structure of Li(Zn,Mn)As and (Ga,Mn)As, all three systems
exhibit exchange interactions and ordered moments of similar
magnitude.

Polycrystalline specimens of (La$_{1-x}$Ba$_x$)(Zn$_{1-x}$Mn$_x$)AsO
were synthesized through the solid state reaction method. High
purity elements of La, Zn, and As were mixed and heated to 900
$^{\circ}$C in an evacuated silica tube to produce intermediate
products LaAs and ZnAs, which were mixed with ZnO, BaO$_2$, and Mn
with nominal concentrations and slowly heated up to 1150
$^{\circ}$C, where the mixture was held for 40 hours before cooling
at 10 $^{\circ}$C/h to room temperature. The polycrystals were
characterized via X-ray diffraction and $dc$-magnetization with a
Quantum Design SQUID. The electrical resistance was measured on
sintered pellets with the typical four-probe method. $\mu$SR
measurements were performed at Paul Scherrer Institute and TRIUMF.

\begin{figure}[!htpb] \centering \vspace*{-2.5cm}
\centering
\includegraphics[width=3.3in]{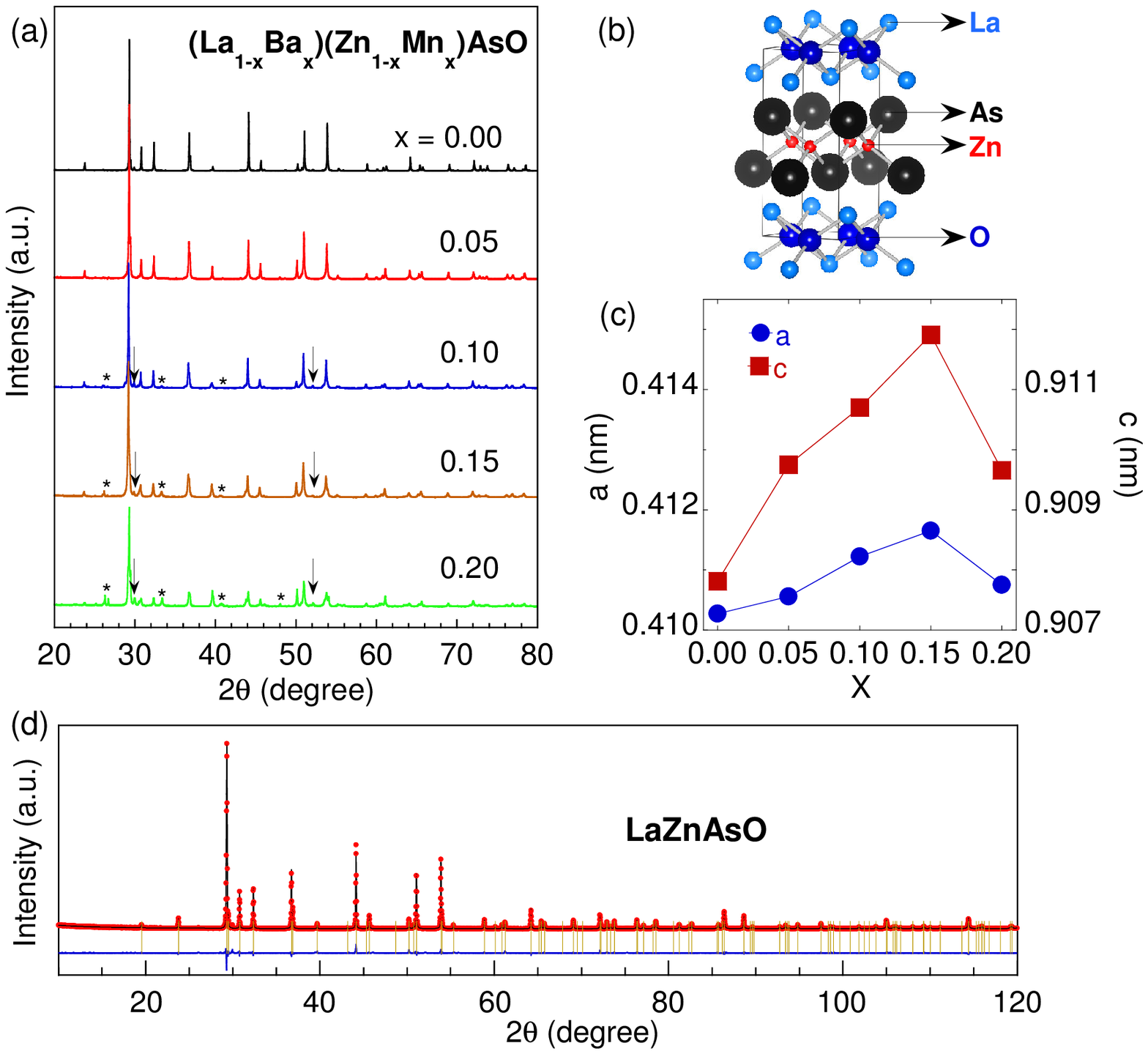}\vspace*{+2cm}
\caption{\label{Fig1:epsart} (Color online). (a) X-ray diffraction
pattern of (La$_{1-x}$Ba$_x$)(Zn$_{1-x}$Mn$_x$)AsO. Traces of
impurity La$_2$O$_3$ ($\downarrow$) and ZnAs$_2$ ($\ast$) are marked
for $x$ $\geq$ 0.1. (b) Crystal Structures of LaZnAsO (P4/nmm). (c)
Lattice constants for the $a$-axis (blue filled circle) and $c$-axis
(red filled square) of (La$_{1-x}$Ba$_x$)(Zn$_{1-x}$Mn$_x$)AsO. (d)
X-ray diffraction pattern of LaZnAsO with Rietveld analyses.}
\end{figure}

The crystal structure and X-ray diffraction patterns are shown in
Fig. 1. Bragg peaks from the parent compound LaZnAsO can be well
indexed by a ZrCuSiAs-type tetragonal crystal structure (space group
P4/nmm), indicating that LaZnAsO is isostructural to LaFeAsO, the
parent compound of the ``1111" family of Fe-based high temperature
superconductors \cite{Kamihara}. The Zn atoms, each one
tetrahedrally coordinated to four As atoms, form parallel square
lattices in the $ab$-plane separated along the $c$-axis by LaO
layers, resulting in the compound's quasi 2D nature. The lattice
parameters were found to be $a$ = 4.1027 {\AA} and $c$ = 9.0781
{\AA}, consistent with the previously reported values $a$ = 4.10492
{\AA} and $c$ = 9.08178 {\AA} \cite{Takano}, and within 5$\%$ of the
LaFeAsO lattice parameters $a$ = 4.0355 {\AA} and $c$ = 8.7393 {\AA}
\cite{Kamihara}. The lattice parameters monotonically increase with
Ba and Mn doping up to $x$ = 0.15, indicating the successful solid
solution of (La,Ba) and (Zn,Mn). Impurity phases of La$_2$O$_3$ and
ZnAs$_2$ start to appear at $x$ = 0.1 and grows prominently through
$x$ = 0.20, as marked by the arrows and stars in Fig. 1(a).

In Fig. 2(b), we show the zero-field cooled (ZFC) and field cooled
(FC) measurements of the $dc$-magnetization $M$ for $B_{ext}$ = 1000
Oe. A significant increase in $M$ is observed at temperatures of 30
K $\sim$ 40 K, and the ZFC and FC curves split at low temperatures
for all doping levels. Interestingly, we do not observe such
features in Mn-doped LaZnAsO, LaZn$_{0.9}$Mn$_{0.9}$AsO, as shown by
the magnetization curve in Fig. 2(a). Instead,
LaZn$_{0.9}$Mn$_{0.9}$AsO remains paramagnetic down to 2 K, where
$M$ is about an order of magnitude smaller than for
(La$_{0.9}$Ba$_{0.1}$)(Zn$_{0.9}$Mn$_{0.1}$)AsO. This indicates that
although doping Mn introduces local moments, the ferromagnetic
ordering will not develop unless the spins are mediated by carriers
arising from (La,Ba) substitutions. This picture is similar to the
case of (Ga,Mn)As system where Zener's model \cite{Dietl2} is
proposed as one candidate to explain the ferromagnetism. In this
theoretical model, RKKY-like interaction of Mn spins are effectively
mediated by hole carriers in the valence band.

\begin{figure}[!htpb] \centering \vspace*{-2cm}
\centering
\includegraphics[width=3.3in]{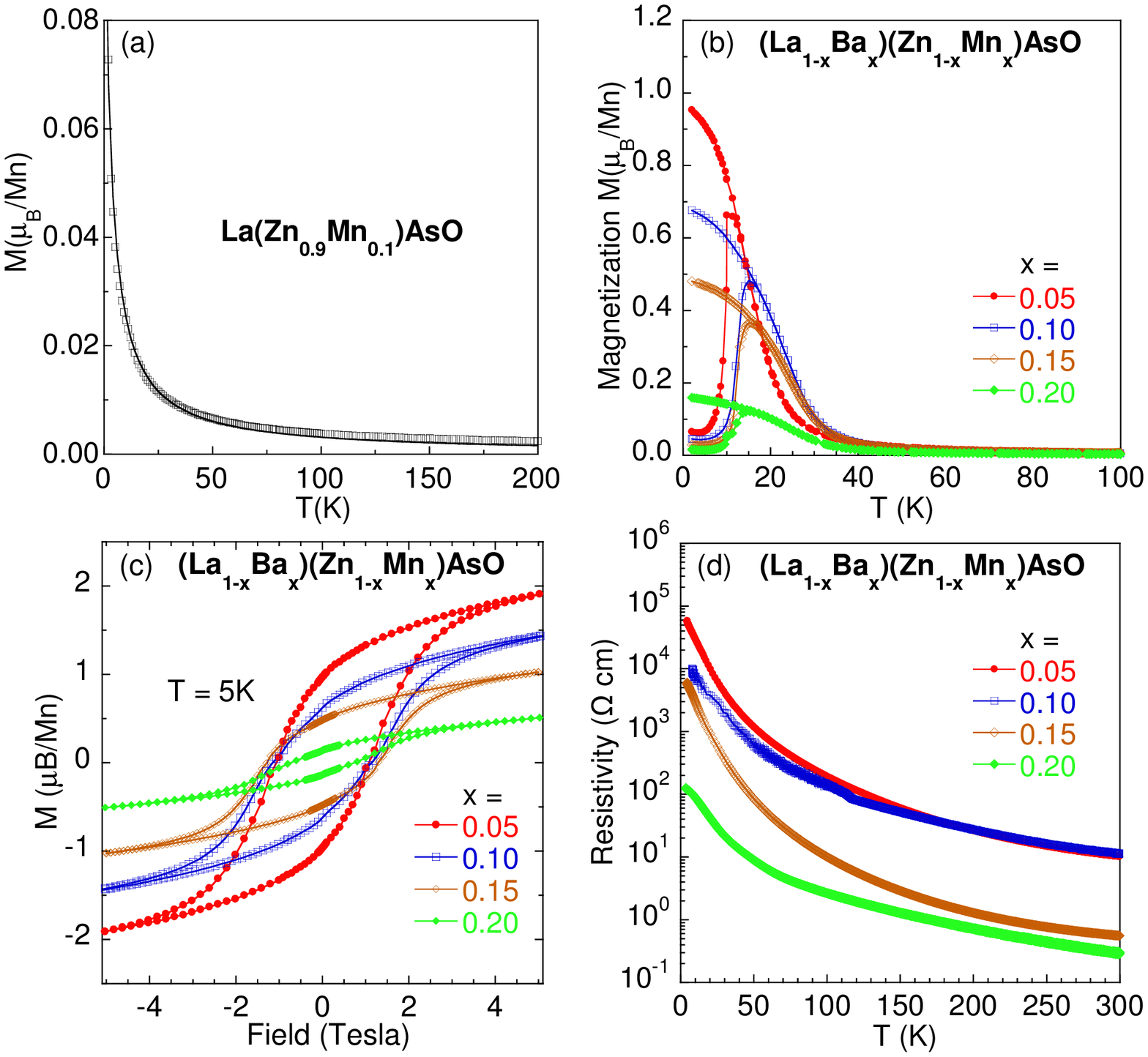}\vspace*{+1.5cm}
\caption{\label{Fig2:epsart} (Color online) (a) The magnetization
$M$ for LaZn$_{0.9}$Mn$_{0.1}$AsO, without charge doping; the solid
line represents the Curie-Weiss law. (b)-(d): Results on
(La$_{1-x}$Ba$_x$)(Zn$_{1-x}$Mn$_x$)AsO: (b) $M$ obtained in the
zero field cooling (ZFC) and field cooling (FC) in the external
field of 1000 Oe. (c) The isothermal magnetization measured at 5 K.
(d) The electrical resistivity.}
\end{figure}

The saturation moment has a maximum of 0.95 $\mu$$_B$/Mn for $x$ =
0.05 and decreases monotonically with increasing $x$, falling to a
value of 0.17 $\mu$$_B$/Mn for $x$ = 0.20. This likely results from
competition between the RKKY interaction, whose first oscillation
period supports ferromagnetic coupling, and nearest neighbor (NN)
antiferromagnetic coupling via direct exchange interaction. For $x$
= 0.1, the probability of finding two Mn atoms at NN Zn sites is
$P(N; x) = C^{4}_{N} \cdot x^{N} \cdot (1-x)^{N}$ = 29.16\%, where N
= 1 and $x$ = 0.1. The direct antiferromagnetic coupling between the
Mn-Mn pairs causes antiferromagnetic ordering in LaMnAsO at $T_N$ =
317 K \cite{Emery}.

We fit the temperature dependence of $M$ above $T_C$ to a
Curie-Weiss law. The effective paramagnetic moment is about 4 - 5
$\mu _B$/Mn, as expected for fully magnetic individual Mn$^{2+}$
moments. The isothermal magnetization of
(La$_{1-x}$Ba$_x$)(Zn$_{1-x}$Mn$_x$)AsO at 5 K is plotted in Fig.
2(c). The parallelogram-shaped hysteresis loops show coercive fields
of 1.06, 1.14, 1.28 and 0.71 T for $x$ = 0.05, 0.10, 0.15 and 0.20,
respectively, much larger than the $\sim$50 Oe coercive field in the
cubic Li$_{1.1}$(Zn$_{0.97}$Mn$_{0.03}$)As \cite{Deng} and
(Ga$_{0.965}$Mn$_{0.035}$)As \cite{Ohno}. The large coercive field
is also reflected in the large differences between the ZFC and FC
curves at low temperature (Fig. 2(b)). The 2D crystal structure of
(La$_{1-x}$Ba$_x$)(Zn$_{1-x}$Mn$_x$)AsO may cause the large coercive
field, as a similar situation was found in (Ba,K)(Zn,Mn)$_2$As$_2$,
another 2D DMS system \cite{Zhao}. Efforts are currently underway to
generate single crystals to further investigate the anisotropic
properties within the $ab$-plane and along the $c$-axis.

In Fig. 2(d), we show electrical resistivity measurements for
(La$_{1-x}$Ba$_x$)(Zn$_{1-x}$Mn$_x$)AsO. All samples display typical
semiconducting behavior over the entire temperature range. The
resistivity decreases as more carriers are introduced through higher
charge doping levels. For $x$ = 0.05, the resistivity is on the
order of 10$^4$ $\Omega$ cm, two orders of magnitude larger than
that of Li(Zn,Mn)As \cite{Deng}. This large resistivity precluded
Hall effect measurements on these polycrystalline specimens. In
recent papers \cite{Liu1,Liu2}, Liu et al. observed Kondo-like
behavior in Mn-doped CaNiGe and CaNiGeH, where the resistivity
decreases linearly with decreasing temperature down to 20 K and then
increases upon further cooling. This is in contrast to the behavior
observed in our compounds.

\begin{figure}[!htpb] \centering \vspace*{-0.1cm}
\centering
\includegraphics[width=3.3in]{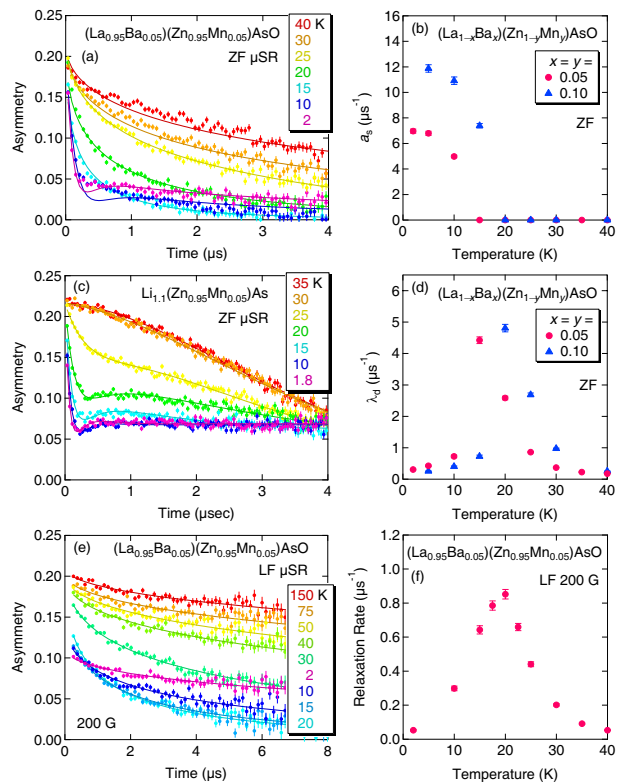}\vspace*{-0.1cm} \\
\caption{\label{Fig3:epsart} (Color online)  Zero field $\mu$SR time
spectra of (a) (La$_{0.95}$Ba$_{0.05}$)(Zn$_{0.95}$Mn$_{0.05}$)AsO
(present work) and (c) Li$_{1.1}$(Zn$_{0.95}$Mn$_{0.05}$)As (adapted
from ref. \cite{Deng}). The solid lines in (a) show a fit to a
dynamic-static relaxation function (eq. 24 of ref. \cite{Uemura})
with the static local field amplitude parameter $a_s$ shown in (b)
and the dynamic relaxation rate parameter $\lambda$$_d$ in (d).  The
solid lines in (c) represent a fit to the ``volume fraction model"
described in ref. \cite{Deng}. (b) and (d) also include the results
for (La$_{0.9}$Ba$_{0.1}$)(Zn$_{0.9}$Mn$_{0.1}$)AsO. Figure (e)
shows the time spectra of LF- $\mu$SR in
(La$_{0.95}$Ba$_{0.05}$)(Zn$_{0.95}$Mn$_{0.05}$)AsO with a
longitudinal field of 200 G and (f) shows the muon spin relaxation
rate 1/T$_1$ due to dynamic spin fluctuations.}
\end{figure}

The availability of bulk specimens enabled us to perform
conventional $\mu$SR experiments on
(La$_{1-x}$Ba$_x$)(Zn$_{1-x}$Mn$_x$)AsO. To determine the nature of
the magnetic order, we conducted zero field (ZF), longitudinal field
(LF), and weak transverse field (WTF) $\mu$SR measurements for $x$ =
0.05 and ZF and WTF measurements for $x$ = 0.1. Fig. 3(a) displays
the time spectra of ZF-$\mu$SR for $x$ = 0.05, showing a rapid
increase of muon spin relaxation for temperatures below $T$ $\sim$
30 K. Several interesting differences are observed between the
current results and the ZF-$\mu$SR time spectra from the cubic
Li$_{1.1}$(Zn$_{0.95}$Mn$_{0.05}$)As, as illustrated in Fig. 3(c)
(adapted from Fig. 3(a) of ref. \cite{Deng}). The spectra for the
``111" system in Fig. 3(c) are well described by the sum of a static
relaxation function representing the magnetically ordered volume and
an exponentially decaying dynamic relaxation function representing
the remaining volume in the paramagnetic phase.  The time spectra of
the present ``1111" system exhibit characteristic signatures of
dynamic slowing down followed by static magnetic order, behavior
also observed in spin glasses AuFe and CuMn \cite{Uemura}. Despite
the imperfections of the fit, evidenced by the differences between
the data and the fit curves in Fig. 3(a), we plot in Figs. 3(b) and
(d) the refined static random field amplitude $a_s$ and dynamic
relaxation rate $\lambda$$_d$ found in the relaxation function given
in eq. 24 of ref. \cite{Uemura}.

To further study the dynamic spin fluctuations, we performed
LF-$\mu$SR measurements on the ``1111" DMS system with $x$ = 0.05 in
LF = 200 G. Analysis of the time spectra displayed in Fig. 3(e)
yields the LF relaxation rate 1/T$_1$ shown in Fig. 3(f), which
exhibits a clear peak at $T$ $\sim$ 15 - 20 K, consistent with the
peak temperature of $\lambda$$_d$ in ZF (Fig. 3(d)) and the onset
temperature of the static spin freezing represented by $a_s$ (Fig.
3(b)). ZF-$\mu$SR results for $x$ = 0.1 yield similar results for
$a_s$ and $\lambda$$_d$, as shown in Figs. 3(b) and (d). LF-$\mu$SR
measurements were not performed on the $x$ = 0.1 system due to
beamtime constraints. We note that for both $x$ = 0.05 and $x$ =
0.1, the onset temperature of $a_s$ in ZF and the ``spin freezing"
temperature indicated by the peaks of $\lambda _d$ in ZF and 1/T$_1$
in LF agree well with the temperature below which the FC and ZFC
magnetization in Fig. 2(b) show a remarkable departure.

In general, the history dependent behavior can be found both in many
regular ferromagnets due to formation and motion of magnetic domains
\cite{Ashcroft}, and in spin glasses due to multiple minima of free
energy as a function of spin configurations \cite{Fischer}. In some
cases z-component of the spin behaves as ferromagnets while x- and
y-components as spin gasses \cite{Fischer, Mirebeau}. Detailed
distinction of these three different cases requires not only the
magnetization data but also neutron scattering results for spatial
spin correlations. Since magnetic neutron scattering signal cannot
be observed due to spatially dilute Mn moments and lack of single
crystal specimens, there is no definite evidence at this moment to
allow distinguishing between ferromagnetic and spin glass states for
the present system.

Practically speaking, however, there is a clear difference between
typical ferromagnets and spin glasses in their magnitudes of the
saturation moment size in the ground state at low temperatures
obtained in zero field after training in high external magnetic
fields. In many ferromagnets, the remnant magnetization value is in
the order of Bohr magneton per magnetic atom, while in typical
dilute alloy spin glasses, it is 0.01 Bohr mangeton per magnetic
atom or less \cite{Tholence,Monod,Prejean}. In the present system,
this remanent magnetization is approximately 1 Bohr magneton per Mn,
as shown in Fig. 2(c).  Therefore, we tentatively assign the present
system to a ferromagnet, rather than to a spin glass.

Use of the so-called ``spin glass relaxation function" \cite{Uemura}
to fit the MuSR data does not provide any distinction between
ferromagnetic and spin glass systems, especially for the cases with
spatially dilute magnetic moments. For example, the earlier MuSR
results on well established ferromagnets (Ga,Mn)As fitted quite well
to the spin glass relaxation function \cite{Dunsiger}.

Analyzed with either the ``volume fraction" fitting used in the
``111" DMS \cite{Deng}, ``122" DMS \cite{Zhao}, and (Ga,Mn)As
\cite{Dunsiger} systems or the ``dynamic spin freezing" model used
in the present ``1111" system, ZF-$\mu$SR results indicate that
these systems all achieve static magnetic order throughout nearly
the entire volume at low temperatures. A closely linear relationship
between the local field amplitude parameter $a_s$ and the Curie
temperature $T_C$ was first noticed in (Ga,Mn)As (Fig. 3(d) of ref.
\cite{Dunsiger}) and Li(Zn,Mn)As (Fig. 3(d) of ref. \cite{Deng}). In
Fig. 4, we plot $a_s$ versus $T_C$ for the two former systems, the
``122" system (Ba,K)(Zn,Mn)$_2$As$_2$ \cite{Zhao}, and the present
(La,Ba)(Zn,Mn)AsO system. The universal linear trend suggests that
the exchange interaction supporting ferromagnetic coupling in these
systems has a common origin and comparable magnitude for a given
spatial density of ordered moments.

\begin{figure}[!htpb] \centering \vspace*{+0.1cm}
\centering
\includegraphics[width=3in]{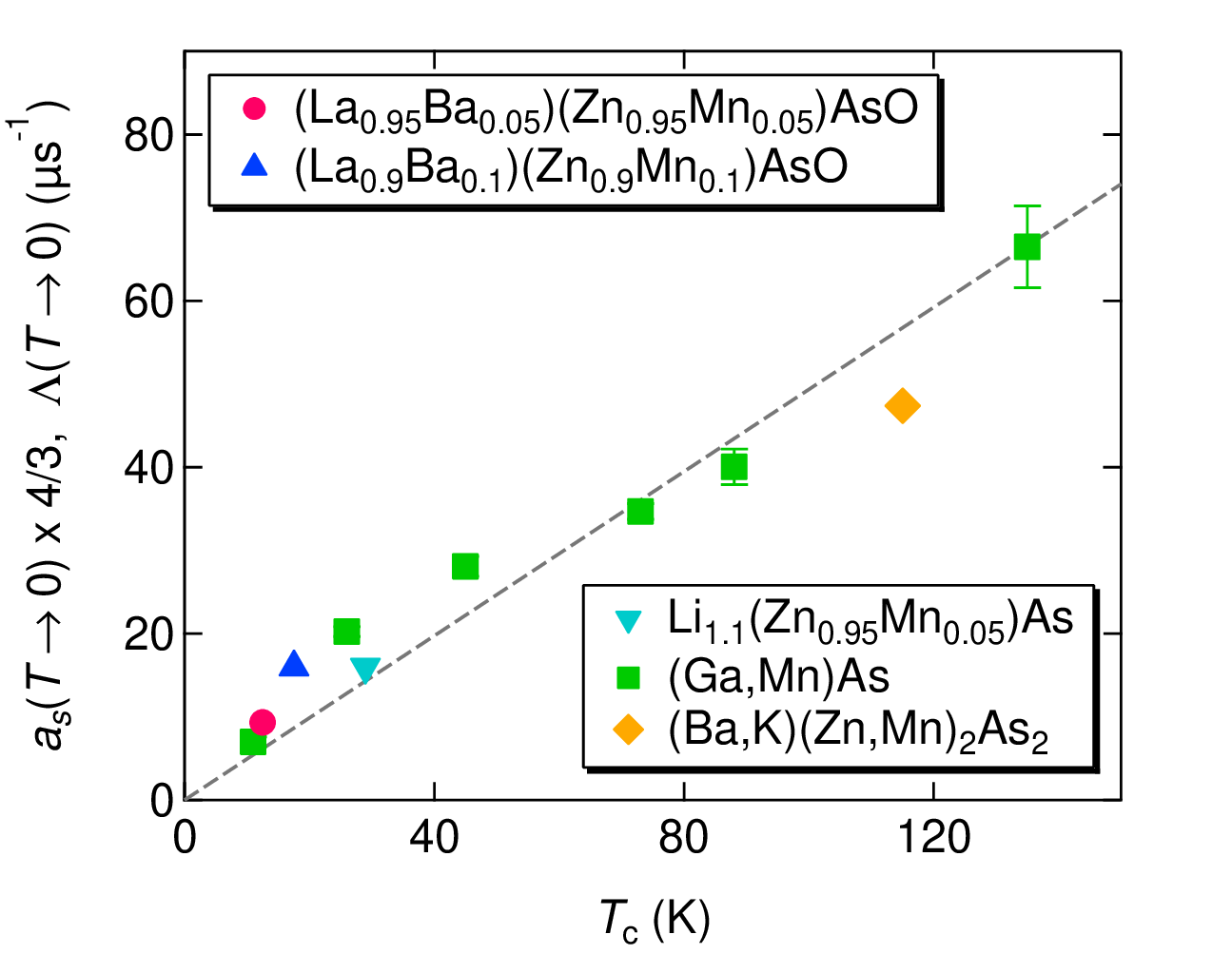}\vspace*{+0.1cm}
\caption{\label{Fig4:epsart} (Color online) Correlation between the
static internal field parameter $a_s$ determined at $T$ = 2 K by
zero-field $\mu$SR versus the ferromagnetic Curie temperature $T_C$
observed in (Ga,Mn)As \cite{Dunsiger}, Li(Zn,Mn)As \cite{Deng},
(La,Ba)(Zn,Mn)AsO (the present work) and (Ba,K)(Zn,Mn)$_2$As$_2$
\cite{Zhao}. Nearly linear correlation indicates a common mechanism
for the ferromagnetic exchange interaction.}
\end{figure}

Magnetization, transport, and $\mu$SR studies carried out in
(Ga$_{1-x}$Mn$_x$)As \cite{Dunsiger} demonstrated that ferromagnetic
order is achieved for small Mn concentrations ($x$ = 0.012-0.030)
even before the system undergoes the semiconductor to metal
transitions. In other words, hole carriers that are not fully
delocalized can mediate the ferromagnetic exchange interaction with
magnitude comparable to those in the case of full delocalization. It
is interesting to note that the quasi 2D ``1111" (present work) and
``122" \cite{Zhao} DMS systems both exhibit ferromagnetism with
relatively high $T_c$ while the charge carriers still remain
semiconducting. The average size of ordered Mn moments in the
``1111" and ``122" systems is significantly smaller than in
(Ga,Mn)As with $T_C$ above $\sim$40 K. This feature suggests that
some of the Mn moments are not involved in the percolating
ferromagnetic network in the semiconducting DMS systems. The
difference between the cubic systems (Ga,Mn)As and Li(Zn,Mn)As and
the 2D DMS systems may also indicate that metallic conduction is
easier to achieve in 3D systems due to a lower percolation
threshold.

In summary, we reported the synthesis of the ``1111" DMS ferromagnet
(La$_{1-x}$Ba$_x$)(Zn$_{1-x}$Mn$_x$)AsO, as the third DMS family
which has a direct counterpart among the FeAs superconductor
families. As discussed in earlier papers \cite{Deng,Zhao}, the
common crystal structure and excellent lattice matching open doors
to the future development of junction devices based on the companion
ferromagnetic and superconducting materials. In parallel with the
present study, the IOP-Beijing group among the present authors has
synthesized another ``1111"  DMS ferromagnet, (La,Ba)(Zn,Mn)SbO,
which will be reported separately \cite{Han}.

The work at Zhejiang University was supported by National Basic
Research Program of China (No. 2011CBA00103), NSFC(No.11274268),
Zhejiang Provincial NSFC(LY12A04006) and Fundamental Research Funds
for Central Universities (2013QNA3016); at IOP in Beijing by the
NSFC and MOST; at Columbia by the US NSF PIRE (Partnership for
International Research and Education: OISE-0968226) and DMR-1105961
projects; the JAEA Reimei project at IOP, Columbia, PSI, McMaster;
and NSERC and CIFAR at McMaster. We would like to thank C. Cao, A.
Fujimori, B. Gu, S.Maekawa and H. Suzuki for helpful discussions.
\\



\begin{thebibliography}{16}
\expandafter\ifx\csname natexlab\endcsname\relax\def\natexlab#1{#1}\fi
\expandafter\ifx\csname bibnamefont\endcsname\relax
  \def\bibnamefont#1{#1}\fi
\expandafter\ifx\csname bibfnamefont\endcsname\relax
  \def\bibfnamefont#1{#1}\fi
\expandafter\ifx\csname citenamefont\endcsname\relax
  \def\citenamefont#1{#1}\fi
\expandafter\ifx\csname url\endcsname\relax
  \def\url#1{\texttt{#1}}\fi
\expandafter\ifx\csname urlprefix\endcsname\relax\def\urlprefix{URL }\fi
\providecommand{\bibinfo}[2]{#2}
\providecommand{\eprint}[2][]{\url{#2}}

\bibitem[{\citenamefont{Ohno et~al.}(1996)\citenamefont{Ohno, Shen,
  Matsukura, Oiwa, Endo, Katsumoto, and Iye}}]{Ohno}
\bibinfo{author}{\bibfnamefont{H.}~\bibnamefont{Ohno}},
  \bibinfo{author}{\bibfnamefont{A.}~\bibnamefont{Shen}},
  \bibinfo{author}{\bibfnamefont{F.}~\bibnamefont{Matsukura}},
  \bibinfo{author}{\bibfnamefont{A.}~\bibnamefont{Oiwa}},
  \bibinfo{author}{\bibfnamefont{A.}~\bibnamefont{Endo}},
  \bibinfo{author}{\bibfnamefont{S.}~\bibnamefont{Katsumoto}}, \bibnamefont{and}
  \bibinfo{author}{\bibfnamefont{Y.}~\bibnamefont{Iye}},
  \bibinfo{journal}{Appl. Phys. Lett.} \textbf{\bibinfo{volume}{69}},
  \bibinfo{pages}{363} (\bibinfo{year}{1996}).

  \bibitem[{\citenamefont{Jungwirth et~al.}(2006)\citenamefont{Jungwirth, Sinova, Masek, Kucera, and MacDonald,}}]{Jungwirth}
\bibinfo{author}{\bibfnamefont{T.}~\bibnamefont{Jungwirth}},
  \bibinfo{author}{\bibfnamefont{J.}~\bibnamefont{Sinova}},
  \bibinfo{author}{\bibfnamefont{J.}~\bibnamefont{Masek}},
  \bibinfo{author}{\bibfnamefont{J.}~\bibnamefont{Kucera}}, \bibnamefont{and}
  \bibinfo{author}{\bibfnamefont{A.H.}~\bibnamefont{MacDonald}},
  \bibinfo{journal}{Rev. Mod. Phys.} \textbf{\bibinfo{volume}{78}},
  \bibinfo{pages}{809} (\bibinfo{year}{2006}).

  \bibitem[{\citenamefont{Dietl et~al.}(2010)\citenamefont{Dietl}}]{Dietl1}
\bibinfo{author}{\bibfnamefont{T.}~\bibnamefont{Dietl}},
  \bibinfo{journal}{Nature Materials} \textbf{\bibinfo{volume}{9}},
  \bibinfo{pages}{965} (\bibinfo{year}{2010}).

  \bibitem[{\citenamefont{Zutic et~al.}(2004)\citenamefont{Zutic, Fabian, and Das Sarma}}]{Zutic}
\bibinfo{author}{\bibfnamefont{I.}~\bibnamefont{Zutic}},
  \bibinfo{author}{\bibfnamefont{J.}~\bibnamefont{Fabian}},\bibnamefont{and}
  \bibinfo{author}{\bibfnamefont{S.}~\bibnamefont{Das Sarma}},
  \bibinfo{journal}{Rev. Mod. Phys.} \textbf{\bibinfo{volume}{76}},
  \bibinfo{pages}{323} (\bibinfo{year}{2004}).

\bibitem[{\citenamefont{Wang et~al.}(2008)\citenamefont{Wang, Campion, Rushforth,
  Edmonds, Foxon, and Gallagher}}]{Wang}
\bibinfo{author}{\bibfnamefont{M.}~\bibnamefont{Wang}},
  \bibinfo{author}{\bibfnamefont{R.P.}~\bibnamefont{Campion}},
  \bibinfo{author}{\bibfnamefont{A.W.}~\bibnamefont{Rushforth}},
  \bibinfo{author}{\bibfnamefont{K.W.}~\bibnamefont{Edmonds}},
  \bibinfo{author}{\bibfnamefont{C.T.}~\bibnamefont{Foxon}}, \bibnamefont{and}
  \bibinfo{author}{\bibfnamefont{B.L.}~\bibnamefont{Gallagher}},
  \bibinfo{journal}{Appl. Phys. Lett.} \textbf{\bibinfo{volume}{93}},
  \bibinfo{pages}{132103} (\bibinfo{year}{2008}).

    \bibitem[{\citenamefont{Samarth et~al.}(2010)\citenamefont{Samarth}}]{Samarth}
\bibinfo{author}{\bibfnamefont{N.}~\bibnamefont{Samarth}},
  \bibinfo{journal}{Nature Materials} \textbf{\bibinfo{volume}{9}},
  \bibinfo{pages}{955} (\bibinfo{year}{2010}).

    \bibitem[{\citenamefont{Chambers et~al.}(2010)\citenamefont{Chambers}}]{Chambers}
\bibinfo{author}{\bibfnamefont{S.}~\bibnamefont{Chambers}},
  \bibinfo{journal}{Nature Materials} \textbf{\bibinfo{volume}{9}},
  \bibinfo{pages}{956} (\bibinfo{year}{2010}).

\bibitem[{\citenamefont{Dunsiger et~al.}(2010)\citenamefont{Dunsiger, Carlo,
  Goko, Nieuwenhuys, Prokscha, Suter, Morenzoni, Chiba, Nishitani, Tanikawa, Matsukura, Ohno, Ohe, Maekawa and Uemura}}]{Dunsiger}
\bibinfo{author}{\bibfnamefont{S.R.}~\bibnamefont{Dunsiger}},
  \bibinfo{author}{\bibfnamefont{J.P.}~\bibnamefont{Carlo}},
  \bibinfo{author}{\bibfnamefont{T.}~\bibnamefont{Goko}},
  \bibinfo{author}{\bibfnamefont{G.}~\bibnamefont{Nieuwenhuys}},
  \bibinfo{author}{\bibfnamefont{T.}~\bibnamefont{Prokscha}},
  \bibinfo{author}{\bibfnamefont{A.}~\bibnamefont{Suter}},
  \bibinfo{author}{\bibfnamefont{E.}~\bibnamefont{Morenzoni}},
  \bibinfo{author}{\bibfnamefont{D.}~\bibnamefont{Chiba}},
  \bibinfo{author}{\bibfnamefont{Y.}~\bibnamefont{Nishitani}},
  \bibinfo{author}{\bibfnamefont{T.}~\bibnamefont{Tanikawa}},
  \bibinfo{author}{\bibfnamefont{F.}~\bibnamefont{Matsukura}},
  \bibinfo{author}{\bibfnamefont{H.}~\bibnamefont{Ohno}},
  \bibinfo{author}{\bibfnamefont{J.}~\bibnamefont{Ohe}},
  \bibinfo{author}{\bibfnamefont{S.}~\bibnamefont{Maekawa}},  \bibnamefont{and}
  \bibinfo{author}{\bibfnamefont{Y.J.}~\bibnamefont{Uemura}},
  \bibinfo{journal}{Nature Materials} \textbf{\bibinfo{volume}{9}},
  \bibinfo{pages}{299} (\bibinfo{year}{2010}).


\bibitem[{\citenamefont{Deng et~al.}(2011)\citenamefont{Deng, Jin,
  Liu, Wang, Zhu, Feng, Chen, Yu, Arguello, Goko, Ning, Zhang, Wang, Aczel, Munsie, Williams, Luke,
  Kakeshita, Uchida, Higemoto, Ito, Gu, Maekawa, Morris and Uemura}}]{Deng}
\bibinfo{author}{\bibfnamefont{Z.}~\bibnamefont{Deng}},
  \bibinfo{author}{\bibfnamefont{C.Q.}~\bibnamefont{Jin}},
  \bibinfo{author}{\bibfnamefont{Q.Q.}~\bibnamefont{Liu}},
  \bibinfo{author}{\bibfnamefont{X.C.}~\bibnamefont{Wang}},
  \bibinfo{author}{\bibfnamefont{J.L.}~\bibnamefont{Zhu}},
  \bibinfo{author}{\bibfnamefont{S.M.}~\bibnamefont{Feng}},
  \bibinfo{author}{\bibfnamefont{L.C.}~\bibnamefont{Chen}},
  \bibinfo{author}{\bibfnamefont{R.C.}~\bibnamefont{Yu}},
  \bibinfo{author}{\bibfnamefont{C.}~\bibnamefont{Arguello}},
  \bibinfo{author}{\bibfnamefont{T.}~\bibnamefont{Goko}},
  \bibinfo{author}{\bibfnamefont{F.L.}~\bibnamefont{Ning}},
  \bibinfo{author}{\bibfnamefont{J.S.}~\bibnamefont{Zhang}},
  \bibinfo{author}{\bibfnamefont{Y.Y.}~\bibnamefont{Wang}},
  \bibinfo{author}{\bibfnamefont{A.A.}~\bibnamefont{Aczel}},
  \bibinfo{author}{\bibfnamefont{T.}~\bibnamefont{Munsie}},
  \bibinfo{author}{\bibfnamefont{T.J.}~\bibnamefont{Williams}},
  \bibinfo{author}{\bibfnamefont{G.M.}~\bibnamefont{Luke}},
  \bibinfo{author}{\bibfnamefont{T.}~\bibnamefont{Kakeshita}},
  \bibinfo{author}{\bibfnamefont{S.}~\bibnamefont{Uchida}},
  \bibinfo{author}{\bibfnamefont{W.}~\bibnamefont{Higemoto}},
  \bibinfo{author}{\bibfnamefont{T.U.}~\bibnamefont{Ito}},
  \bibinfo{author}{\bibfnamefont{B.}~\bibnamefont{Gu}},
  \bibinfo{author}{\bibfnamefont{S.}~\bibnamefont{Maekawa}},
  \bibinfo{author}{\bibfnamefont{G.D.}~\bibnamefont{Morris}},  \bibnamefont{and}
  \bibinfo{author}{\bibfnamefont{Y.J.}~\bibnamefont{Uemura}},
  \bibinfo{journal}{Nature Communications} \textbf{\bibinfo{volume}{2}},
  \bibinfo{pages}{422} (\bibinfo{year}{2011}).


  \bibitem[{\citenamefont{Masek et~al.}(2007)\citenamefont{Masek, Kudrnovsky, Maca, Gallagher, Campion, Gregory, and Jungwirth}}]{Masek}
\bibinfo{author}{\bibfnamefont{J.}~\bibnamefont{Masek}},
  \bibinfo{author}{\bibfnamefont{J.}~\bibnamefont{Kudrnovsky}},
  \bibinfo{author}{\bibfnamefont{F.}~\bibnamefont{Maca}},
  \bibinfo{author}{\bibfnamefont{B.L.}~\bibnamefont{Gallagher}},
  \bibinfo{author}{\bibfnamefont{R.P.}~\bibnamefont{Campion}},
  \bibinfo{author}{\bibfnamefont{D.H.}~\bibnamefont{Gregory}}, \bibnamefont{and}
  \bibinfo{author}{\bibfnamefont{T.}~\bibnamefont{Jungwirth}},
  \bibinfo{journal}{Phys. Rev. Lett.} \textbf{\bibinfo{volume}{98}},
  \bibinfo{pages}{067202} (\bibinfo{year}{2007}).

\bibitem[{\citenamefont{Wang et~al.}(2011)\citenamefont{Wang, Liu, Lv, Gao, Yang, Yu, Li, Jin}}]{Wangxc}
\bibinfo{author}{\bibfnamefont{X.C. }~\bibnamefont{Wang}},
  \bibinfo{author}{\bibfnamefont{Q.Q.}~\bibnamefont{Liu}},
  \bibinfo{author}{\bibfnamefont{Y.Y.}~\bibnamefont{Lv}},
  \bibinfo{author}{\bibfnamefont{W.B.}~\bibnamefont{Gao}},
  \bibinfo{author}{\bibfnamefont{S.M.}~\bibnamefont{Feng}},
  \bibinfo{author}{\bibfnamefont{L.X.}~\bibnamefont{Yang}},
  \bibinfo{author}{\bibfnamefont{R.C.}~\bibnamefont{Yu}},
  \bibinfo{author}{\bibfnamefont{F.Y.}~\bibnamefont{Li}}, \bibnamefont{and}
 \bibinfo{author}{\bibfnamefont{C.Q.}~\bibnamefont{Jin}},
  \bibinfo{journal}{Solid State Communications} \textbf{\bibinfo{volume}{148}},
  \bibinfo{pages}{538} (\bibinfo{year}{2008}).

\bibitem[{\citenamefont{Parker et~al.}(2011)\citenamefont{Parker, Pitcher, Baker, Franke, Lancaster, Blundell, Clarke}}]{Parker}
\bibinfo{author}{\bibfnamefont{D.R. }~\bibnamefont{Parker}},
  \bibinfo{author}{\bibfnamefont{M.J.}~\bibnamefont{Pitcher}},
  \bibinfo{author}{\bibfnamefont{P.J.}~\bibnamefont{Baker}},
  \bibinfo{author}{\bibfnamefont{I.}~\bibnamefont{Franke}},
  \bibinfo{author}{\bibfnamefont{T.}~\bibnamefont{Lancaster}},
  \bibinfo{author}{\bibfnamefont{S.J.}~\bibnamefont{Blundell}}, \bibnamefont{and}
 \bibinfo{author}{\bibfnamefont{S.J.}~\bibnamefont{Clarke}},
  \bibinfo{journal}{Chemical Communications},
  \bibinfo{pages}{2189} (\bibinfo{year}{2009}).

  \bibitem[{\citenamefont{Zhao et~al.}(2013)\citenamefont{Zhao, Deng, Wang, Han, Zhu, Li, Liu, Goko, Frandsen, Liu, Ning, Uemura, Dabkovska,
  Luke, Leutkens, Molenzoni, Dunsiger, Senyshyn, Boeni, and Jin}}]{Zhao}
\bibinfo{author}{\bibfnamefont{K.}~\bibnamefont{Zhao}},
  \bibinfo{author}{\bibfnamefont{Z.}~\bibnamefont{Deng}},
  \bibinfo{author}{\bibfnamefont{X.C.}~\bibnamefont{Wang}},
  \bibinfo{author}{\bibfnamefont{W.}~\bibnamefont{Han}},
  \bibinfo{author}{\bibfnamefont{J.L.}~\bibnamefont{Zhu}},
  \bibinfo{author}{\bibfnamefont{X.}~\bibnamefont{Li}},
  \bibinfo{author}{\bibfnamefont{Q.Q.}~\bibnamefont{Liu}},
  \bibinfo{author}{\bibfnamefont{T.}~\bibnamefont{Goko}},
  \bibinfo{author}{\bibfnamefont{B.}~\bibnamefont{Frandsen}},
  \bibinfo{author}{\bibfnamefont{L.}~\bibnamefont{Liu}},
  \bibinfo{author}{\bibfnamefont{F.L.}~\bibnamefont{Ning}},
  \bibinfo{author}{\bibfnamefont{Y.J.}~\bibnamefont{Uemura}},
  \bibinfo{author}{\bibfnamefont{H.}~\bibnamefont{Dabkovska}},
  \bibinfo{author}{\bibfnamefont{G.M.}~\bibnamefont{Luke}},
  \bibinfo{author}{\bibfnamefont{H.}~\bibnamefont{Leutkens}},
  \bibinfo{author}{\bibfnamefont{E.}~\bibnamefont{Molenzoni}},
  \bibinfo{author}{\bibfnamefont{S.R.}~\bibnamefont{Dunsiger}},
  \bibinfo{author}{\bibfnamefont{A.}~\bibnamefont{Senyshyn}},
  \bibinfo{author}{\bibfnamefont{P.}~\bibnamefont{Boeni}}, \bibnamefont{and}
  \bibinfo{author}{\bibfnamefont{C.Q.}~\bibnamefont{Jin}},
  \bibinfo{journal}{Nature Communications} \textbf{\bibinfo{volume}{4}},
  \bibinfo{pages}{1422} (\bibinfo{year}{2013}).

  \bibitem[{\citenamefont{Johnston et~al.}(2006)\citenamefont{Johnston,}}]{Johnston}
\bibinfo{author}{\bibfnamefont{D.C}~\bibnamefont{Johnston}},
   \bibinfo{journal}{Adv. Phys.} \textbf{\bibinfo{volume}{59}},
  \bibinfo{pages}{803} (\bibinfo{year}{2010}).

\bibitem[{\citenamefont{Kamihara et~al.}(2008)\citenamefont{Kamihara, Watanabe,
  Hirano, and Hosono}}]{Kamihara}
\bibinfo{author}{\bibfnamefont{Y.}~\bibnamefont{Kamihara}},
  \bibinfo{author}{\bibfnamefont{T.}~\bibnamefont{Watanabe}},
  \bibinfo{author}{\bibfnamefont{M.}~\bibnamefont{Hirano}}, \bibnamefont{and}
  \bibinfo{author}{\bibfnamefont{H.}~\bibnamefont{Hosono}},
  \bibinfo{journal}{J. Am. Chem. Soc.} \textbf{\bibinfo{volume}{130}},
  \bibinfo{pages}{3296} (\bibinfo{year}{2008}).

\bibitem[{\citenamefont{Kayanuma et~al.}(2009)\citenamefont{Kayanuma, Kawamura,
  and Hosono}}]{Kayanuma}
\bibinfo{author}{\bibfnamefont{K.}~\bibnamefont{Kayanuma}},
  \bibinfo{author}{\bibfnamefont{R.}~\bibnamefont{Kawamura}},
 \bibinfo{author}{\bibfnamefont{H.}~\bibnamefont{Hiramatsu}},
  \bibinfo{author}{\bibfnamefont{H.}~\bibnamefont{Yanagi}},
   \bibinfo{author}{\bibfnamefont{M.}~\bibnamefont{Hirano}},
    \bibinfo{author}{\bibfnamefont{T.}~\bibnamefont{Kamiya}},
  \bibnamefont{and}
  \bibinfo{author}{\bibfnamefont{H.}~\bibnamefont{Hosono}},
  \bibinfo{journal}{Thin Solid Films} \textbf{\bibinfo{volume}{516}},
  \bibinfo{pages}{5800} (\bibinfo{year}{2008}).


\bibitem[{\citenamefont{Takano et~al.}(2008)\citenamefont{Takano, Komaatsuzaki,
  Komasaki, Watanabe, Takahashi, and Takase}}]{Takano}
\bibinfo{author}{\bibfnamefont{Y.}~\bibnamefont{Takano}},
  \bibinfo{author}{\bibfnamefont{S.}~\bibnamefont{Komaatsuzaki}},
  \bibinfo{author}{\bibfnamefont{H.}~\bibnamefont{Komasaki}},
  \bibinfo{author}{\bibfnamefont{T.}~\bibnamefont{Watanabe}},
  \bibinfo{author}{\bibfnamefont{Y.}~\bibnamefont{Takahashi}}, \bibnamefont{and}
  \bibinfo{author}{\bibfnamefont{K.}~\bibnamefont{Takase}},
  \bibinfo{journal}{J. Alloys. Compd.} \textbf{\bibinfo{volume}{451}},
  \bibinfo{pages}{467} (\bibinfo{year}{2008}).

\bibitem[{\citenamefont{Dietl et~al.}(2000)\citenamefont{Dietl, Ohno,
  Matsukura, Cibert, and Ferrand}}]{Dietl2}
\bibinfo{author}{\bibfnamefont{T.}~\bibnamefont{Dietl}},
  \bibinfo{author}{\bibfnamefont{H.}~\bibnamefont{Ohno}},
  \bibinfo{author}{\bibfnamefont{F.}~\bibnamefont{Matsukura}},
  \bibinfo{author}{\bibfnamefont{J.}~\bibnamefont{Cibert}}, \bibnamefont{and}
  \bibinfo{author}{\bibfnamefont{D.}~\bibnamefont{Ferrand}},
  \bibinfo{journal}{Science} \textbf{\bibinfo{volume}{287}},
  \bibinfo{pages}{1019} (\bibinfo{year}{2000}).

\bibitem[{\citenamefont{Emery et~al.}(2011)\citenamefont{Emery, Wildman, Skakle, Mclaughlin, Smith and Fitch}}]{Emery}
\bibinfo{author}{\bibfnamefont{N.}~\bibnamefont{Emery}},
  \bibinfo{author}{\bibfnamefont{E.J.}~\bibnamefont{Wildman}},
  \bibinfo{author}{\bibfnamefont{J.M.S.}~\bibnamefont{Skakle}},
  \bibinfo{author}{\bibfnamefont{A.C.}~\bibnamefont{Mclaughlin}},
  \bibinfo{author}{\bibfnamefont{R.I.}~\bibnamefont{Smith}},  \bibnamefont{and}
  \bibinfo{author}{\bibfnamefont{A.N.}~\bibnamefont{Fitch}},
  \bibinfo{journal}{Phy. Rev. B} \textbf{\bibinfo{volume}{83}},
  \bibinfo{pages}{144429} (\bibinfo{year}{2011}).

\bibitem[{\citenamefont{Liu et~al.}(2011)\citenamefont{Liu, Matsuishi,
  Fujitsu, and Hosono}}]{Liu1}
\bibinfo{author}{\bibfnamefont{X.F.}~\bibnamefont{Liu}},
  \bibinfo{author}{\bibfnamefont{S.}~\bibnamefont{Matsuishi}},
  \bibinfo{author}{\bibfnamefont{S.}~\bibnamefont{Fujitsu}}, \bibnamefont{and}
  \bibinfo{author}{\bibfnamefont{H.}~\bibnamefont{Hosono}},
  \bibinfo{journal}{Phys. Rev. B} \textbf{\bibinfo{volume}{84}},
  \bibinfo{pages}{214439} (\bibinfo{year}{2011}).

\bibitem[{\citenamefont{Liu et~al.}(2012)\citenamefont{Liu, Matsuishi,
  Fujitsu, Ishigaki, Kamiyama, and Hosono}}]{Liu2}
\bibinfo{author}{\bibfnamefont{X.F.}~\bibnamefont{Liu}},
  \bibinfo{author}{\bibfnamefont{S.}~\bibnamefont{Matsuishi}},
  \bibinfo{author}{\bibfnamefont{S.}~\bibnamefont{Fujitsu}},
  \bibinfo{author}{\bibfnamefont{T.}~\bibnamefont{Ishigaki}},
  \bibinfo{author}{\bibfnamefont{T.}~\bibnamefont{Kamiyama}}, \bibnamefont{and}
  \bibinfo{author}{\bibfnamefont{H.}~\bibnamefont{Hosono}},
  \bibinfo{journal}{J. Am. Chem. Soc.} \textbf{\bibinfo{volume}{134}},
  \bibinfo{pages}{11687} (\bibinfo{year}{2012}).

\bibitem[{\citenamefont{Uemura et~al.}(1985)\citenamefont{Uemura, Yamazaki,
  Harshman, Senba, and Ansaldo}}]{Uemura}
\bibinfo{author}{\bibfnamefont{Y.J.}~\bibnamefont{Uemura}},
  \bibinfo{author}{\bibfnamefont{T.}~\bibnamefont{Yamazaki}},
  \bibinfo{author}{\bibfnamefont{D.R.}~\bibnamefont{Harshman}},
  \bibinfo{author}{\bibfnamefont{M.}~\bibnamefont{Senba}}, \bibnamefont{and}
  \bibinfo{author}{\bibfnamefont{E.J.}~\bibnamefont{Ansaldo}},
  \bibinfo{journal}{Phys. Rew. B} \textbf{\bibinfo{volume}{31}},
  \bibinfo{pages}{546} (\bibinfo{year}{1985}).


  \bibitem[{\citenamefont{Ashcroft}(2010)\citenamefont{Ashcroft}}]{Ashcroft}
\bibinfo{author}{\bibfnamefont{N.W.}~\bibnamefont{Ashcroft}},
  \bibnamefont{and}
\bibinfo{author}{\bibfnamefont{N.D.}~\bibnamefont{Mermin}},
\bibinfo{journal}{\textit{Solid State Physics}}
  (\bibinfo{journal}{Holt, Rinehart and Winston, 1976}).


  \bibitem[{\citenamefont{Fischer}(2010)\citenamefont{Fischer}}]{Fischer}
\bibinfo{author}{\bibfnamefont{K.H.}~\bibnamefont{Fischer}},
  \bibnamefont{and}
\bibinfo{author}{\bibfnamefont{J.A.}~\bibnamefont{Hertz}},
\bibinfo{journal}{\textit{Spin Glasses}}
  (\bibinfo{journal}{Cambridge University Press, 1991}).


  \bibitem[{\citenamefont{Mirebeau}(1992)\citenamefont{Mirebeau}}]{Mirebeau}
\bibinfo{author}{\bibfnamefont{I.}~\bibnamefont{Mirebeau}},
\bibinfo{pages}{P41,}
\bibinfo{journal}{\textit{Recent Progress in Random Magnets}}
  (\bibinfo{journal}{Edited by D.H. Ryan, World Scientific Publishing Co. Ptc. Ltd, 1992}).

\bibitem[{\citenamefont{Tholence.}(1985)\citenamefont{Tholence}}]{Tholence}
\bibinfo{author}{\bibfnamefont{J.L.}~\bibnamefont{Tholence}},
 \bibnamefont{and}
  \bibinfo{author}{\bibfnamefont{R.}~\bibnamefont{Tournier}},
  \bibinfo{journal}{J. Phys. (Paris)} \textbf{\bibinfo{volume}{35}},
  \bibinfo{pages}{C4-229} (\bibinfo{year}{1974}).

\bibitem[{\citenamefont{Prejean.}(1985)\citenamefont{Prejean}}]{Prejean}
\bibinfo{author}{\bibfnamefont{J.J.}~\bibnamefont{Prejean}},
\bibinfo{author}{\bibfnamefont{M.}~\bibnamefont{Joliclerc}},
 \bibnamefont{and}
  \bibinfo{author}{\bibfnamefont{P.}~\bibnamefont{Monod}},
  \bibinfo{journal}{J. Phys. (Paris)} \textbf{\bibinfo{volume}{41}},
  \bibinfo{pages}{427} (\bibinfo{year}{1980}).

\bibitem[{\citenamefont{Monod}(1985)\citenamefont{Monod}}]{Monod}
\bibinfo{author}{\bibfnamefont{P.}~\bibnamefont{Monod}},
\bibinfo{author}{\bibfnamefont{J.J.}~\bibnamefont{Prejean}},
 \bibnamefont{and}
  \bibinfo{author}{\bibfnamefont{B.}~\bibnamefont{Tissier}},
  \bibinfo{journal}{J. Appl. Physics} \textbf{\bibinfo{volume}{50}},
  \bibinfo{pages}{7324} (\bibinfo{year}{1979}).

  \bibitem[{\citenamefont{Han et~al.}(2013)\citenamefont{Han, Deng, Wang, Han, Zhu, Li, Liu, Goko, Frandsen, Liu, Ning, Uemura, Dabkovska,
  Luke, Leutkens, Molenzoni, Dunsiger, Senyshyn, Boeni, and Jin}}]{Han}
\bibinfo{author}{\bibfnamefont{W.}~\bibnamefont{Han}},
  \bibinfo{author}{\bibfnamefont{F.L.}~\bibnamefont{Ning}},
  \bibinfo{author}{\bibfnamefont{Y.J.}~\bibnamefont{Uemura}},
  \bibnamefont{and}
  \bibinfo{author}{\bibfnamefont{C.Q.}~\bibnamefont{Jin}},
  \bibinfo{journal}{to be submitted}.

\end{thebibliography}

\end{document}